\documentclass[fleqn,usenatbib]{mnras}

\usepackage{newtxtext,newtxmath}
\usepackage{cleveref}
\usepackage[T1]{fontenc}

\DeclareRobustCommand{\VAN}[3]{#2}
\let\VANthebibliography\thebibliography
\def\thebibliography{\DeclareRobustCommand{\VAN}[3]{##3}\VANthebibliography}

\usepackage{soul}
\usepackage[dvipsnames]{xcolor}

\usepackage{graphicx}	% Including figure files
\usepackage{amsmath}	% Advanced maths commands
\usepackage{xcolor}		% textset part of text in color for easy todos
\usepackage{physics}	% nice package for physics equations
\usepackage{multicol}   % Extra package for the multi-figure
\usepackage{nicefrac}
\usepackage[switch]{lineno}
\usepackage{csquotes}
\title[Two Colour Intensity Interferometry H.E.S.S.]{Simultaneous Two Colour Intensity Interferometry with H.E.S.S.}

\author[N Vogel et al.]{
Naomi Vogel$^{1}$\thanks{E-mail: naomi.vogel@fau.de},
Andreas Zmija$^{1}$,
Frederik Wohlleben$^{1,2}$,
Gisela Anton$^{1}$,
Alison Mitchell$^{1}$,
\newauthor Adrian Zink$^{1}$, Stefan Funk$^{1}$
\\
$^{1}$Erlangen Centre for Astroparticle Physics, Friedrich-Alexander-Universität Erlangen-Nürnberg, Nikolaus-Fiebiger-Str. 2, Erlangen 91058, Germany\\
$^{2}$now at Max-Planck-Institut für Kernphysik, Saupfercheckweg 1, Heidelberg, 69117, Germany\\
}

\date{Accepted XXX. Received YYY; in original form ZZZ}

\pubyear{\the\year{2024}}

\begin{document}
\label{firstpage}
\pagerange{\pageref{firstpage}--\pageref{lastpage}}
\maketitle

\begin{abstract}
In recent years, intensity interferometry has been successfully applied to the Imaging Atmospheric Cherenkov Telescopes H.E.S.S.\,, MAGIC, and VERITAS. All three telescope systems have proven the feasibility and capability of this method. After our first campaign in 2022, when two of the H.E.S.S.\ telescopes in Namibia were equipped with our external setup and the angular diameter of two stars was measured, our setup was upgraded for a second campaign in 2023, where the goal is to perform simultaneous two colour measurements. The second campaign not only involves a third equipped telescope, but also each mechanical setup now includes two interference filters at two different wavelengths ($375\,$nm and $470\,$nm) with a broader bandwidth of $10\,$nm. This enables having simultaneous two-colour measurements, which yields information about the star's physical size at different wavelengths. This is the first time that simultaneous dual-waveband intensity interferometry measurements are performed. The angular diameter results of the 4 stars, Mimosa ($\beta$ Cru), Eta Centauri ($\eta$ Cen), Nunki ($\sigma$ Sgr) and Dschubba ($\delta$ Sco), are reported, where the effects of limb darkening are also taken into account.
\end{abstract}


%%%%%%%%%%%%%%%%%%%%%%%%%%%%%%%%%%%%%%%%%%%%%%%%%%

%%%%%%%%%%%%%%%%% BODY OF PAPER %%%%%%%%%%%%%%%%%%

%#####################
\section{Introduction}
%#####################

Intensity interferometry (II) was first employed in the 1960s by Robert Brown and Richard Q. Twiss (HBT), resulting in a successful operation of the Narrabri Stellar Intensity Interferometer that measured the angular diameter of 32 stars \citep{brown1967stellar, HBT_32}. As the development of fast electronics and large area collectors was lagging behind, the technique was discarded in the field of astronomy to make way for stellar amplitude interferometry which has developed into notable interferometric instrumentations, such as CHARA \citep{CHARA_description} and the VLTI \citep{haguenauer2010very}. The technique exploits the first order coherence of electromagnetic waves from a source observed by separated telescopes, connecting the phases and amplitudes. Due to atmospheric turbulence the baselines between the telescopes are limited and thus the optical resolution to some $0.1\,$mas. In contrast to amplitude interferometry, intensity interferometry exploits the second order correlation, the square of the wave amplitude relating to its intensity, making it essentially unresponsive to either the telescope's optical deficiencies or to atmospheric turbulence. The possibility of kilometer-long baselines in the optical wavelength range raises the concept of microarcsecond imaging. \citep{dravins2016intensity} 

Therefore, intensity interferometry has gained additional astronomical interest in the last two decades and efforts have been made to implement it using Imaging Atmospheric Cherenkov Telescopes (IACTs), which are ground based gamma-ray telescopes. Their advantages are the large diameters ($>10\,$m) and the array-like layout not only allowing for baselines on the order of $100\,$m \citep{LeBohec2006} up to $2\,$km but also offering many different baselines at the same time in the case of the future Cherenkov Telescope Array Observatory (CTAO) \citep{Dravins2013}. 

The IACT observatories, H.E.S.S., MAGIC, and VERITAS have already successfully demonstrated the efficiency of realising stellar intensity interferometry (SII) measurements. The VERITAS-SII measured the diameters of two B stars ($\beta$ CMa and $\varepsilon$ Ori) with all its four telescopes \citep{abeysekara2020demonstration} and is extending their measurements to stars of higher magnitude and making further improvements concerning the angular resolution \citep{kieda2022performance}. Their recent publication is about their angular diameter measurements of the star $\beta\,$ UMa \citep{acharyya2024} at visual wavelengths ($416\,$nm), being consistent with the results of CHARA measurements in the infrared region. MAGIC started out with measuring the correlation signal at different telescope baselines of three stars, namely $\varepsilon$ CMa (Adhara), $\eta$ UMa (Benetnasch) and $\beta$ CMa (Mirzam) \citep{acciari2020optical}. Their active mirror control allows for sub-telescope baselines and zero-baseline correlation measurements. An improved sensitivity is expected by adding the CTA LST-1 to the interferometer \citep{cortina2022first}. They recently published measurements of 22 stellar diameters \citep{Abe_2024}. 

In April 2022 first intensity interferometry measurements were conducted with the H.E.S.S.\ telescopes in Namibia by mounting II setups to the lid of the camera of two telescopes. Two single stars in the southern hemisphere, Shaula ($\lambda$ Sco) and Nunki ($\sigma$ Sgr), were successfully observed. Due to the installation of a beamsplitter in the optics of each telescope zero-baseline correlations were possible. \citep{Zmija_2023}

After having proven that the approach of II works with the H.E.S.S.\ telescopes, our team considered alternative concepts for interferometric application, starting with simultaneous two-colour II measurements. This paper presents our second II campaign in April/May 2023 at the H.E.S.S.\ telescopes where we not only integrated a third telescope into our interferometer, but also expanded our optical setup to two different colour filters ($375\,$nm and $470\,$nm). This is the first time that II measurements were performed simultaneously in two different wavebands at an IACT and also represent the shortest wavelength that has been employed thus far for II observations. Three single stars in the southern hemisphere were observed and also studied by HBT, Mimosa ($\beta$ Cru), Eta Centauri ($\eta$ Cen) and Dschubba ($\delta$ Sco) \citep{HBT_32}. In addition we repeated measurements on Nunki ($\sigma$ Sgr) for comparison with two colours, as well as reducing statistical uncertainties despite having less observation time (see Sect.~\ref{Nunki}). In summary, this demonstrates that multi-waveband II measurements are feasible and could provide the potential to enhance signal-to-noise in future II observations.

Hereafter, we first review the physical quantities of SII in Sect.~\ref{sec:observables}. The differences of the experimental setup with its new application is described in Sect.~\ref{sec:meas_setup}. In Sect.~\ref{sec:meas_prod} we briefly revisit our measurement procedure and in Sect.~\ref{sec:analysis} outline the details of our analysis for the recorded data. Section \ref{sec:results} gathers the final angular diameters of our target stars.

%##########################
\section{Theoretical Review}\label{sec:observables}
%##########################
Here we provide a short theoretical review (see \cite{HanburyBrown1974} for further information).

The projected distance between two telescopes as seen from a given star is the so called projected baseline $r$, which changes when the star moves along the sky. An intensity interferometer correlates the intensity signals $I_1$ and $I_2$ recorded at each telescope as a function of the projected baseline, resulting in the second order correlation function

\begin{equation}
g^{(2)}(r, \tau) = \frac{\expval{I_1(0,t) I_2(r, t + \tau)}}{\expval{I_1} \expval{I_2}}\,,
\end{equation}

where $\tau$ is an applied time shift between the two signals and $\expval{}$ the average over time $t$. 

The Siegert relation \citep{siegert1943fluctuations, Siegert_Ferreira} connects the second order correlation function to the first order correlation function by
\begin{equation}
g^{(2)}(r, \tau) = 1 + |g^{(1)}(r, \tau)|^{2}\,,
\end{equation}

and the first order correlation can also be separated into a spatial and temporal part
\begin{equation}\label{3}
g^{(1)}(r, \tau) =g^{(1)}(r) g^{(1)}(\tau)\,.
\end{equation}

The spatial component, also referred to as visibility, is expressed by the Fourier transform of the source's spatial intensity distribution, which is described by the van Cittert-Zernike theorem \citep{Mandel1995}. Especially for a star that is considered as a disc of uniform intensity up to an angular diameter $\theta$, the first order correlation is given by

\begin{equation}\label{spcoh}
g^{(1)}(r) = 2 \frac{J_1(\pi r \theta / \lambda)}{ \pi r \theta / \lambda }\,,
\end{equation}

where $J_1$ is the Bessel function of first kind and $\lambda$ the wavelength of light. The second order correlation $ g^{(2)}(r) - 1 = |g^{(1)}(r)|^{2} $\,, observed in an intensity interferometer, is referred to as squared visibility. The so called resolving baseline is identified by the first zero at $r \approx 1.22 \lambda/\theta$ and increases for smaller sources or larger wavelengths. Measuring the squared visibility over a broad range of baselines (squared visibility curves) allows us to determine the angular size of the star.
\\
The temporal component in Eq.~\ref{3} is expressed by the Fourier transform of the light spectral density which is described by the Wiener-Khinchin theorem \citep{Mandel1995}. In a measurement of $g^{(2)}(\tau)$ a photon bunching peak should be visible in the temporal range of the coherence time

\begin{equation}\label{tauc}
    \tau_c := \frac{\lambda_0^2}{c\Delta\lambda}, 
\end{equation}

with $c$ being the speed of light, $\lambda_0$ the central wavelength and $\Delta\lambda$ the optical bandwidth. When using optical filters with a bandwidth of the order $\Delta\lambda = 1$\,nm the coherence time is of order $1$\,ps, which is much shorter than what can be resolved by detector systems. Instead, only the integral under $ g^{(2)}(r) - 1$ can be measured. As the value of the coherence time decreases to broader bandwidths, but photon rates and thus photon statistics increase, the signal-to-noise of a measurement is independent of the used optical bandwidth.

%##########################
\section{The SII system at H.E.S.S.}\label{sec:meas_setup}
% system, synonym for setup 
%##########################
\subsection{The H.E.S.S. Telescopes}

\begin{figure}
    \centering
    \includegraphics[width=\columnwidth]{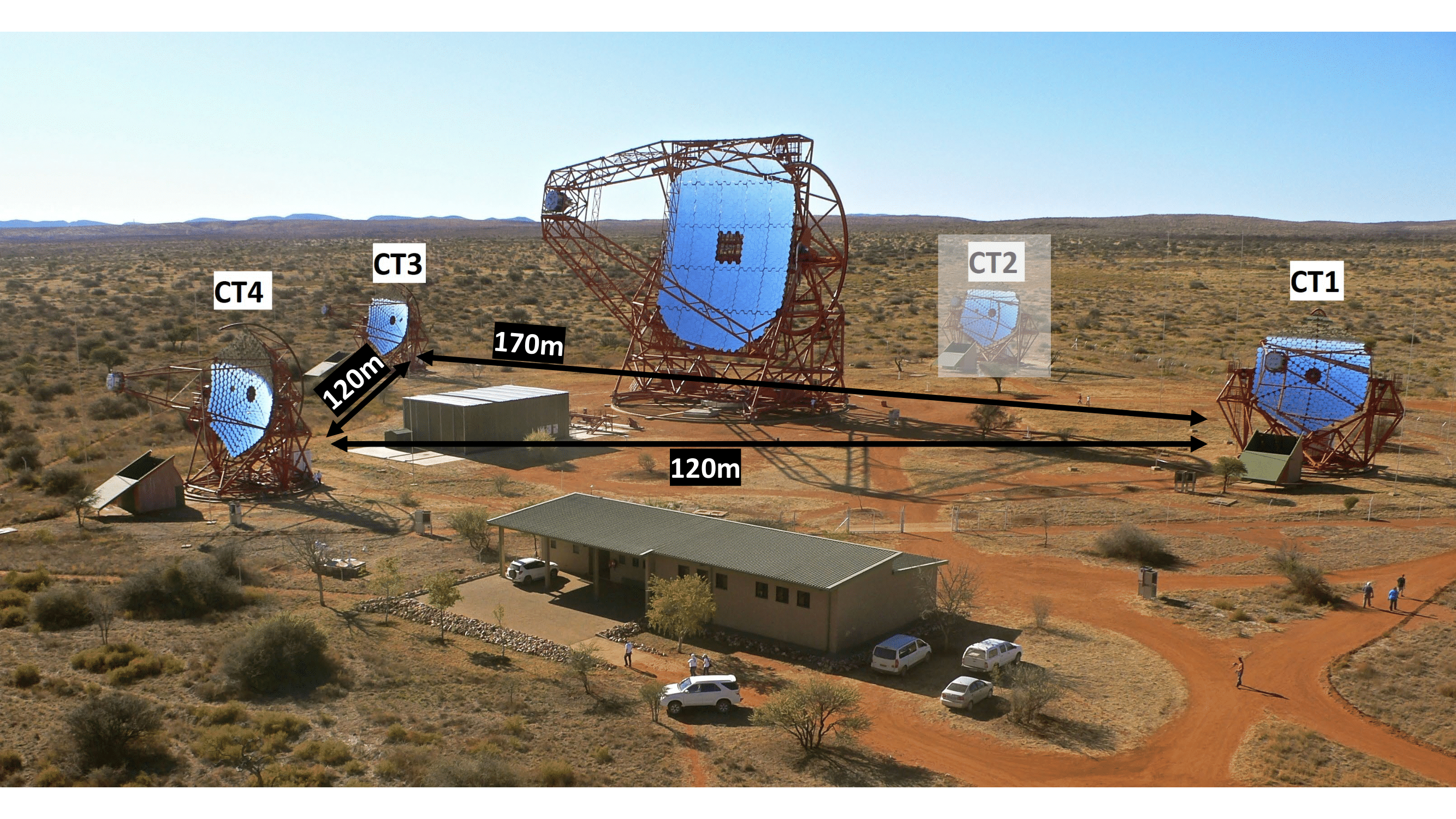}
    \caption{The H.E.S.S. site. In 2002 the Phase I telescopes (CT1-4) were inaugurated. The central telescope was built in 2012. In 2023 the intensity interferometry setups were mounted on telescopes CT1, CT3 and CT4. The control building (bottom centre) is used to operate the telescopes and for part of the data analysis. \protect\citep{HESSpic}}
    \label{fig:hess}
\end{figure}

For the intensity interferometry campaign in 2023 three of the H.E.S.S.\ telescopes (Phase I) were used, which are situated in the Khomas Highlands in Namibia at an altitude of 1835 m.a.s.l.. Four IACTs make up a square with $120\,$m side length and $170\,$m diagonal length, with a 600\,m${^2}$ mirror area telescope at its centre, which was added in 2012 (see Fig.~\ref{fig:hess}). The mirror area of a phase I telescope is compromised of 382 round mirror facets leading to a total area of 108\,m${^2}$, a diameter of $12\,$m and a focal length of $15\,$m. The telescopes are built according to the Davis-Cotton design. \citep{HESStel} The main goal for intensity interferometry is a sub-milliarcsecond resolution which is reached by relating large light collecting areas and long telescope baselines. These requirements make the H.E.S.S.\ telescopes suitable candidates. 

\subsection{Experimental Setup}\label{sec:exp}

\begin{figure}
    \centering
    \includegraphics[width=\columnwidth]{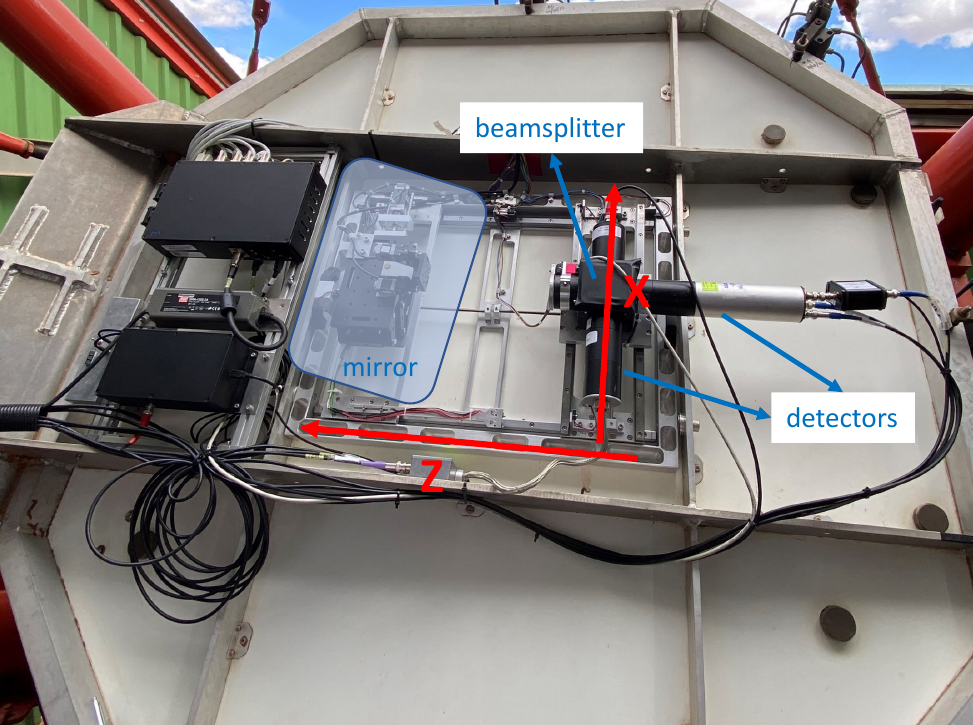}
    \caption{Intensity interferometry setup mounted to the lid of the telescope CT3. The mirror was not installed when the picture was taken. Its position is indicated by the blue shaded area on the left side of the setup. The red arrows mark the range of motion for each individual component, which is equipped with motors.}
    \label{fig:setup}
\end{figure}

\begin{figure}
    \includegraphics[width=\columnwidth]{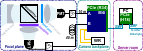}
    \caption{Schematic of the intensity interferometer including optics and data transfer. Each telescope has an identical setup. The blue bandpass filter for the transmitted light has a wavelength of $470\,$nm and the purple one for the reflected beam a wavelength of $375\,$nm. Both have a bandwidth of $10\,$nm. The main difference to the optics of the 2022 campaign is the order of lenses and filters. For the single colour measurements in 2022 the order was \textit{lens-filter-lens-beamsplitter-detectors}. For the two waveband measurements in 2023 the order was \textit{lens-beamsplitter-filters-lens-detectors}.}
    \label{fig:setup_schematic}
\end{figure}

The II setup was designed to fit on the lid of the H.E.S.S.\ Phase I cameras. Figure \ref{fig:setup} shows the setup mounted in the centre of the camera lid in the focal spot of the H.E.S.S.\ mirrors. The setup is described in \cite{Zmija_2023}. A third setup was constructed according to this design. The rectangular base ($50 \times 63$)\,cm${^2}$ is made of aluminium with a total weight of $21\,$kg including all attached equipment. Figure \ref{fig:setup_schematic} shows a schematic of the setup. The $45^\circ$ angled mirror reflects the incoming light beam (from the left side) into the optical system and onto the two detectors.  \\
The mirror for the additional third telescope is also a SEA-VIS front-surface mirror produced via CNC precision-cut by Präzisions Glas \& Optik GmbH and is glued to an aluminum frame for stabilisation. \\
Since the measurements for the 2023 campaign are done with two colour filters simultaneously (see Fig.~\ref{fig:setup_schematic}) the optics within the optical path are arranged differently than for the previous one colour measurements. The light beam first encounters a parallelizing lens (coated concave lens, $F=-75$\,mm) in a 2" tube system and is then split into two beams by a dichroic beamsplitter plate with a cut-on wavelength at $415\,$nm. The transmitted beam is lead through a LC-HBP470/10-50 narrowband optical (interference) filter (470\,nm central wavelength CWL, 10\,nm width) and the reflected beam through a LC-HBP375/10-50 narrowband optical (interference) filter (375\,nm CWL, 10\,nm width). Both beams are then focused onto the active area of the photo-multiplier tube (PMT) using a converging lens (coated bi convex lens, $F=100$\,mm). The lenses are manufactured by \citep{Thorlabs}, the beamsplitter and the interference filter by \citep{Lasercomponents}. We decided to increase the bandwidth of the interference filters from $2\,$nm to $10\,$nm after successful laboratory experiments with broader optical bandwidths. We confirmed that we are able to control higher photon rates than the ones we had in the 2022 campaign and also measure smaller correlation peaks. The reason for the increase in bandwidth is to decrease systematic errors in the measurements. In particular, the aim is to decrease the uncertainty on the actual effective filter bandwidth, which comes from light beams hitting the interference filters at non-perpendicular inclinations, which are always present due to imperfect parallelisation of the collimated beam.
Since we use two interference filters with different CWL, we decided to also use two different photomultiplier types manufactured by Hamamatsu for optimal performance. For the filter with $470\,$nm we use the PMT R11265U-300 \citep{Hamamatsu} with a peak sensitivity at $420\,$nm and a quantum efficiency of 39\%. It has a quantum efficiency of 30\% at the relevant wavelength of $470\,$nm and was also used in our campaign in 2022. For the filter with $375\,$nm we use the PMT R11265U-200 \citep{Hamamatsu} with a peak sensitivity at $400\,$nm and a quantum efficiency of 43\% also at $375\,$nm. The High Voltage (HV) power supply is placed in the camera backplane and supplies the first 9 dynodes. To stabalize the voltage at high rates, the last 4 dynodes are powered by a further booster power supply. The amplifier is situated directly behind the PMT on the right side (transmitted beam) and on the bottom of the setup for the lower PMT (reflected beam) to reduce electronic noise (see Fig.~\ref{fig:setup}). The amplified PMT signals are transported through airborne 5 low-loss, double shielded coaxial cables with $10\,$m and $40\,$m length to evade any crosstalk at the sampling interface.\\  
To counteract any inevitable mis-pointing of the telescopes which for H.E.S.S.\ operation is not online-corrected, each item of the setup is assembled with stepper motors, which can be moved remotely, to assure that a maximum amount of the light beam is captured by the optical system. The mis-pointing is on the order of $<1\,$cm - $5\,$cm depending on the elevation. The red arrows in Fig.~\ref{fig:setup} demonstrate the range of motion for the different items of the system. The optics and detectors can be moved in the camera plane (x- and z-axis) and the mirror in z-axis and y-axis (up and down). The motors rotating the mirror around the x- and z-axis were removed since they didn't show any sufficient improvement in bringing the spot more into focus during the campaign in 2022. The motor control box and its power supply are positioned to the left of the setup on the lid. Both HV cables and signal cables are brought to the camera backplane where the HV power supply and the digitizer are installed. \\ 
Besides the HV power supply a M4i.2211-x8 digitizer from \cite{specint} is installed in two of the telescopes' backplane racks as described in \cite{Zmija_2023}. The four channel version of this digitizer is installed in the third telescope, although we only use two of them for the two PMTs. After the PMT currents are transformed to voltages, which are proportional to the currents, they are sampled by the digitizers at a speed of $625\,$MS/s at a digitization range of $\pm200\,$mV. Using the R34 PCIe board system from \cite{pcie}, the PCIe bus of the workstation is extended to each telescope via optical fiber.\\
It is important that all three data streams are synchronized. This is ensured by connecting each digitizer to the White Rabbit (WR) network of the H.E.S.S.\ telescopes. The pulse-per-second (PPS) of the WR triggers the digitizers and they sample 2 GS of data synchronously, which amounts to $3.436\,$seconds of data taking. The next sampling starts with the PPS 4 seconds after the previous triggered sampling and the process is repeated. The duty cycle of the procedure is $85.9\%$. A $10\,$MHz clock is used as reference to generate a synchronized sampling timebase for all three telescope digitizers. \\
During the measurements the data are stored on hard disks to be correlated offline during the day or after the campaign.

%#################################################
\section{Measurement procedure}\label{sec:meas_prod}
%#################################################
As gamma ray observations don't take place during the H.E.S.S.\ moonlight break -- when the moon has a fraction of lunar illumination of 0.6 and is above the horizon -- we can make use of that time with our intensity interferometry observations, which were carried out in April/May 2023. This campaign's main targets of observations are the stars Mimosa ($\beta$ Crucis), Eta Centauri, Dschubba ($\delta$ Scorpii) and, for comparison to the campaign of 2022, Nunki ($\sigma$ Sgr). They were selected taking into account their configuration, trajectory, magnitude and separation to the moon. All results are presented in Sect.~\ref{sec:results}.\\
The procedure of our measurements is identical to that of the 2022 campaign described in \cite{Zmija_2023}. At the beginning of our measurement time for one night, the three telescopes are positioned on one of the chosen stars and start tracking it until targets are changed (When the star is less than $\approx 30^\circ$ above the horizon) or the night is over (when the sun is only $10^\circ$ below the horizon). The second step is to perform a PMT calibration with the HV turned off in order to extract the direct current (DC) offset in the analysis later on. Then a PMT pulse calibration with the HV turned on is done to establish the live photon rates by using the shape and height distribution of the photon pulses. The incoming photon rates have to be low for calibration, which we accomplished by moving our motor system so our optics are just out of focus of the star. Details of the calibration can be found in \cite{10.1093/mnras/stab3058}.\\
To focus the optics directly on the star, we perform our online pointing correction by adjusting the individual parts of the setup with the motors to achieve maximal photon rates. When the optimal photon rate is found, the data taking begins. Due to mis-pointing of the telescopes it happens that over time during the observations the optical system is no longer in the focus of the star beam (as noticed via a drop in photon rate) and the optical elements are adjusted via moving the motors. When the measurements end the HV is shut down and the motors are moved to their parking position. The telescopes are parked, the amplifiers turned off and the correlation analysis of the acquired data starts.

%############################################################
\section{Data Analysis}\label{sec:analysis}
%############################################################
\subsection{Correlation procedure and data corrections}
The temporal correlation of the PMT currents(waveforms) is performed after the measurements are taken. As we use three telescopes (CT1, CT3 and CT4), each with two colours (channel A and B) (see Sect.~\ref{sec:meas_setup}), there are three binary files, each including the two channels A and B. Using the correlation function from the python package \textit{cupy}, the six waveform channels are first read from the hard disk and then correlated. For each colour channel there are three possible telescope combinations (1-3, 1-4 and 3-4). Due to a malfunctioning amplifier not all three telescopes were in operation simultaneously the whole time but only two at a time. Therefore, the stars Mimosa and Eta Centauri only show the combinations 1-4 and 3-4 in later analysis results. At the end of the campaign all three telescopes were running simultaneously, seen in the Nunki and Dschubba results. 

Before further analysis is done, the correlated data are cleaned by adding a lowpass filter with a cut-off frequency of $200\,$MHz and notch filters to remove sharp noise frequencies. Furthermore, the optical path delay correction is applied to correct for the difference of the optical path length at the different telescopes. More detailed information about the noise cuts and the optical path delay correction can be found in \cite{Zmija_2023}. Since only channels with similar cable lengths between the telescopes, e.g. CT1 channel A with CT4 channel A (both $\approx 10\,$m cable length) or CT1 channel B with CT4 channel B (both $\approx 40\,$m cable length), are correlated no cable delay correction has to be applied.

\subsection{Data preparation and signal extraction}
\begin{figure}
    \includegraphics[width=\columnwidth]{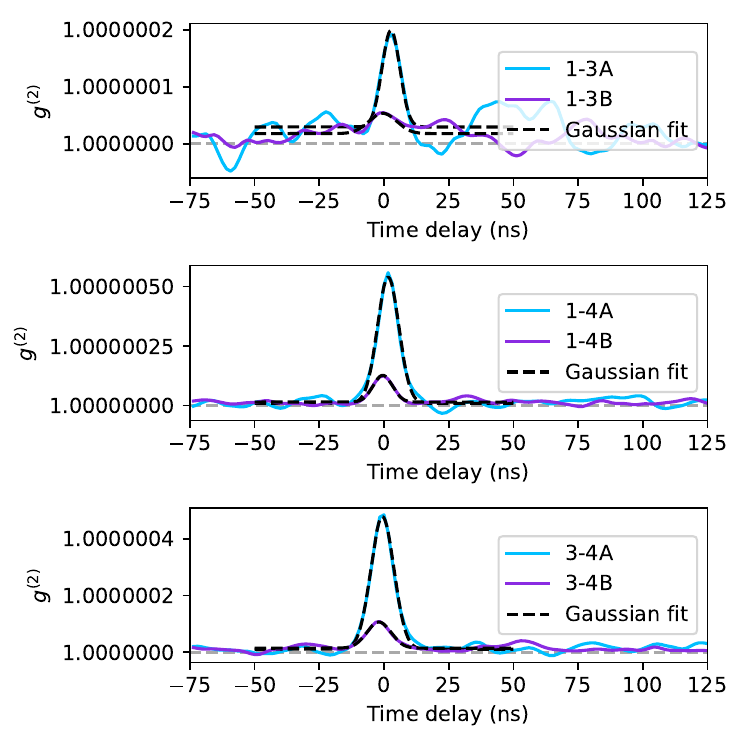}
    \caption{$g^{(2)}$ functions of the accumulated data of all stars for each telescope combination. The upper plot shows telescope combination 1-3, the middle 1-4 and the lower 3-4. The blue curve represents the $g^{(2)}$ function of the measurements of channel A with the 470\,nm interference filter and the purple curve the measurements of channel B with the 375\,nm filter. The black dashed line is the Gaussian fit (including vertical offset) to each data curve.}
    \label{fig:crosscorr}
\end{figure}

After the data corrections are applied, the data of each star is bundled into time segments to track the change in squared visibility and determine the angular size of the star. For each segment the temporal $g^{(2)}$ function is calculated by normalization of the correlated data. However, when splitting the data into segments it becomes apparent that the photon bunching peak is not always significantly detected, especially at large baselines. This complicates fitting a Gaussian function to the signal peak to extract the squared visibility. To stabilize the fit in each segment global Gaussian fit parameters are extracted. Therefore, the data of all stars for each telescope combination is collected and averaged, with each data array weighted by the inverse square of the standard error (RMS) of the noise of the $g^{(2)}$ function in order to account for the different data quality. The averaged $g^{(2)}$ functions are shown in Fig.~\ref{fig:crosscorr}. For each telescope combination there is a $g^{(2)}$ function for channel A shown in blue and for channel B in purple. A Gaussian function is fitted to each signal peak implied with the black dashed line. The correlation signal is always located at $\tau \sim 0$ therefore fixing the mean of the Gaussian function is not necessary. The amplitudes of the $g^{(2)}$ functions for both colours for the telescope combination 1-3 are smaller than for the other combinations due to 1-3 having the largest projected baselines during the measurements. The channel B data ($375\,$nm, purple) always show smaller amplitudes than the channel A data ($470\,$nm, blue). This is due to two factors: the coherence time $\tau_c \approx \lambda_0^2/(c\Delta\lambda)$ being smaller for light at smaller wavelengths, and the squared visibility curves for smaller wavelengths decreasing more rapidly with increasing baselines (see Eq.~\ref{spcoh}), meaning taking an average squared visibility over a range of baselines results in smaller values for smaller wavelengths.\\
For each combination and channel the width of the Gaussian (sigma) is extracted (see Fig.~\ref{fig:crosscorr}). To receive a global sigma for each channel the weighted average of the sigma value of the telescope combinations is calculated. We determine a global sigma of $4.33\,$ns for channel A and $4.06\,$ns for channel B. 

\subsection{Squared visibility curves}
\begin{figure}
    \includegraphics[width=\columnwidth]{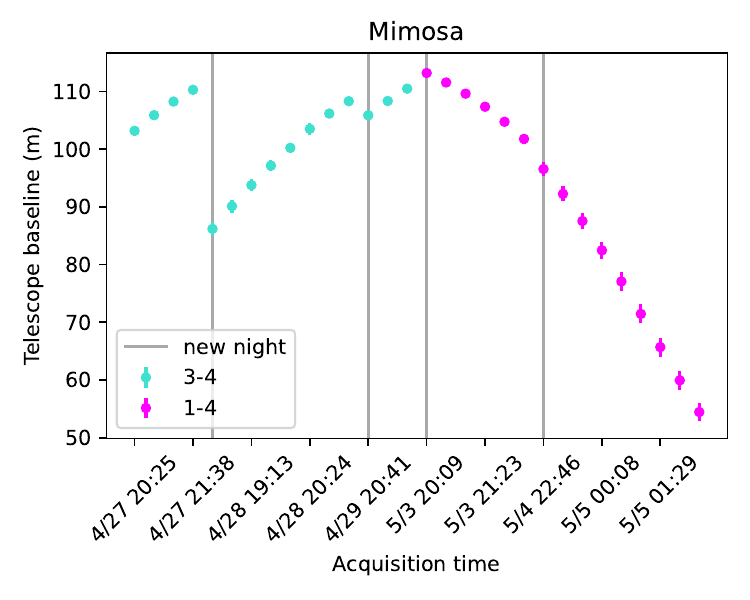}
    \caption{Telescope baselines for each measurement segment of the star Mimosa. The vertical gray lines indicate a new measurement night. The different colours represent the telescope combinations with which the measurement was taken. This plot shows the change of the projected baseline over the whole measurement period of Mimosa.}
    \label{fig:baselines}
\end{figure}

For accurate analysis of the change in squared visibility the data of each star is divided into time segments. The length of these segments depends on the brightness of the star and the acquisition time. The details are described in \cite{Zmija_2023}. Table \ref{table:measurement_chunks} shows an overview of the segments and measurement times of each star. As the star moves on its trajectory during the night the projected baseline changes. This is shown in Fig.~\ref{fig:baselines} where the projected baseline is plotted against the acquisition time of each segment over the complete measurement time of Mimosa. The corresponding baseline for each time segment is derived by taking the average of the baselines of each data file in one segment. The errorbars imply the range of baselines within one measurement segment and are the $1\sigma$ distribution. The cyan points represent measurements taken with telescope combination 3-4 and the pink points with telescope combination 1-4. The vertical gray lines indicate a new observation night. 

To extract the squared visibility a Gaussian is fitted to the $g^{(2)}$ function of each segment. In order to prevent fitting into fluctuations we fix the Gaussian width sigma, which is quite stable. Thus, we only fit the mean and the amplitude of the Gaussian. The integral of the fitted Gaussian is plotted against the projected baseline, as seen in Fig.~\ref{fig:sc_mimosa_etacen} for each colour separately. Again, the cyan points represent measurements taken with telescope combination 3-4 and the pink points with telescope combination 1-4. In order to properly consider effects of the stochastic photon noise on the uncertainty, we derive the errors of the integral with the following approach: the photon bunching peak, modelled as a Gaussian, is extracted by a fit into the data, and then added onto 160 independent intervals of the $g^{(2)}$ function left and right of the actual peak, which usually consist of different realizations of stochastic noise only. In each interval the peak is re-fitted and its integral computed. The $1\,\sigma$ distribution of the reconstructed integrals are considered the uncertainty on the measurement. 

\begin{table}
\caption{List of the measurement segments and segment times in minutes for each star.}
\centering
\begin{tabular}{ c  c  c  c  c } 
 \hline
 \noalign{\smallskip}
 \parbox[c]{1cm}{Night \\ (mm/dd)} & Mimosa & Eta Centauri & Nunki & Dschubba \\ [0.5ex] 
 \hline\hline 
 \noalign{\smallskip}
 04/27 & $4 \times 24.13$ & -                     & -                     & -                     \\
 04/28 & $8 \times 22.20$ & -                     & -                     & -                     \\
 04/29 & $3 \times 25.40$ & -                     & -                     & -                     \\
 05/03 & $6 \times 24.67$ & -                     & -                     & -                     \\
 05/04 & $9 \times 27.27$ & -                     & -                     & -                     \\
 05/06 & -                & -                     & $8 \times 40.00$      & -                     \\
 05/07 & -                & $14 \times 41.74$     & -                     & -                     \\
 05/08 & -                & $ 8 \times 45.40$     & -                     & -                     \\
 05/09 & -                & -                     & -                     & $6 \times 37.27$      \\
 05/10 & -                & -                     & $3 \times 40.80$      & $6 \times 34.93$      \\
 \hline
 \noalign{\smallskip}
 \textbf{Total} & 12.4h & 15.79h & 7.37h & 7.22h \\
 \hline
\end{tabular}
\label{table:measurement_chunks}
\end{table}

\subsection{Angular diameter analysis}
To analyse the angular diameter of our targets, we take both the uniform disc (UD) and the limb darkened (LD) model into account. In the case of the UD model (see Eq.~\ref{spcoh}) we use the following fit function:

\begin{equation}
f(b) = A \cdot |g^{(1)}(b)|^2 = A \left[ \frac{2J_1(\pi b \theta_{\mathrm{UD}} / \lambda)}{ \pi b \theta_{\text{UD}} / \lambda }\right]^2 \,,
\end{equation}

where $b$ is the projected baseline, and $\lambda$ the central wavelength of the interference filter (either $470\,$nm or $375\,$nm). For fitting we use the orthogonal distance regression (odr) algorithm from \textit{scipy}, which takes both vertical and horizontal errorbars into account. From the fit we gather the angular diameter $\theta_{\text{UD}}$ and the "zero baseline amplitude" $A$. The expected value of the latter, $A_{\text{expect}}$, corresponds to

\begin{equation}
    A_{\text{expect}} = \int_{-\infty}^{+\infty} \left( g^{(2)}_{\text{expect}}(\tau, b=0) -1 \right) d\tau = 0.5 \cdot k_T \cdot \tau_c
\end{equation}

 where $\tau_c$ is defined in Eq.~\ref{tauc} and the factor $k_T$ is a calculated correction factor, which depends on the profile of the interference filters. The details are described in \cite{Zmija_2023} and the results are $0.83$ for the $375\,$nm filter and $0.84$ for the $470\,$nm filter. Since our light is unpolarized a factor of $0.5$ is added to the equation. This leads to an expected zero baseline amplitude of $A_{\text{expect}} = 19.45\,$fs for the $375\,$nm filter and $A_{\text{expect}} = 31.09\,$fs for the $470\,$nm filter, whereas the fit gives a parameter which is smaller by a factor of $2$ and $1.6$ respectively. Table \ref{table:ZB} shows the fitted zero baseline amplitudes for our targets and different filter colours. One potential reason for the loss in coherence is a remaining bandwidth mismatch of the interference filters between the telescopes, see Sect.~\ref{sec:exp}. An indication for this is that the loss is less severe compared to the measurements at $2\,$nm bandwidth in 2022. But the exact origin is still unclear and being investigated. The correlation at zero baseline is a fixed parameter of our instruments, not depending on parameters of an observed star. Therefore, we take the zero baseline amplitudes of our investigated targets computed via the UD fit model and estimate the weighted average for each colour. The result is also noted in Table \ref{table:ZB}. Next, we insert this value as a data point, seen e.g. in Fig.~\ref{fig:sc_mimosa_etacen} as black data point at $b=0$, and refit the UD model to all data. The resulting angular diameter of the second fit is the main outcome we document for each star and colour. 

As mentioned above, we also account for limb darkening (LD) by applying the following equation introduced by \cite{HanburyBrown1974}:

\begin{align}
f(b) = A \bigg[ \frac{1-u_{\lambda}}{2} + \frac{u_{\lambda}}{3} \bigg]^{-2} \bigg[ (1-u_{\lambda}) & \frac{J_1(x_{\text{LD}})}{x_{\text{LD}}} \\
&+ u_{\lambda} \sqrt{\pi/2} \frac{J_{3/2}(x_{\text{LD}})}{(x_{\text{LD}})^{3/2}} \Bigg]^2 \,, \notag
\end{align}

where $u_{\lambda}$ is the limb darkening coefficient, $x_{\text{LD}} = \pi b \theta_{\text{LD}} / \lambda $ and $\theta_{\text{LD}}$ is the limb darkened angular diameter. The interpolation of the limb darkening coefficient $u_{\lambda}$ is done as described in \cite{abeysekara2020demonstration} with look-up tables from \cite{LD}.

\begin{table*}
\caption{Zero baseline amplitudes computed by the UD fit model in fs for each filter colour and their weighted average. }
\centering
\begin{tabular}{ c  c  c  c  c } 
 \hline
 \noalign{\smallskip}
Filter & Mimosa & \parbox[c]{1cm}{Eta \\ Centauri} & Nunki & \parbox[c]{1cm}{weighted \\ average} \\ [0.1ex] 
\noalign{\smallskip}
 \hline\hline 
 \noalign{\smallskip}
 $470\,$nm & $19.80 \pm 1.86$ & $22.68 \pm 2.44$ & $15.89 \pm 2.13$ & $19.23 \pm 1.22$  \\
 $375\,$nm & $12.06 \pm 1.89$ & $ 8.60 \pm 1.51$ & $ 6.61 \pm 2.83$ & $ 9.45 \pm 1.09$  \\
 \hline
\end{tabular}
\label{table:ZB}
\end{table*}

%############################################################
\section{Results}\label{sec:results}
%############################################################
In this section, we summarize the results of our measurement campaign in 2023. During our observation period data was taken of four stars in the southern hemisphere. Mimosa ($\beta$ Crux) and Eta Centauri ($\eta$ Cen) are both single stars and were chosen as they were also investigated by HBT, which makes them good candidates for comparison. Nunki ($\sigma$ Sgr) is the third single star which was observed in 2023. We studied Nunki in 2022 \citep{Zmija_2023} and will compare those results in Sect.~\ref{Nunki}. The last star, which was on our target list, was Dschubba ($\delta$ Sco), a spectroscopic binary. We will not present a binary analysis in this paper but use a single star model to fit to the data, which is clarified in the next section. Dschubba was also observed by HBT and treated as a single star \citep{HBT_32}.

\subsection{Stellar diameter results}
\begin{figure}
    \includegraphics[width=\columnwidth]{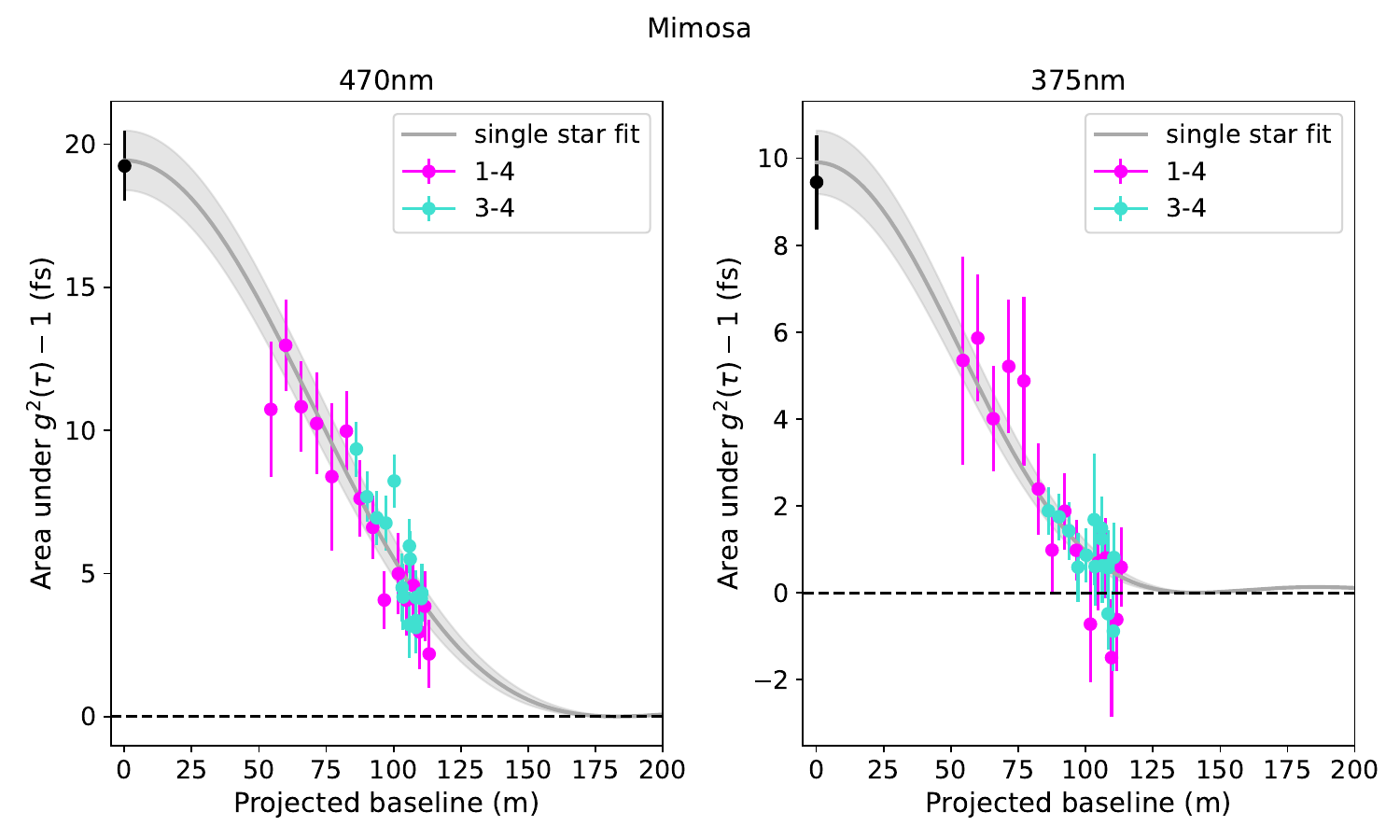}
    \includegraphics[width=\columnwidth]{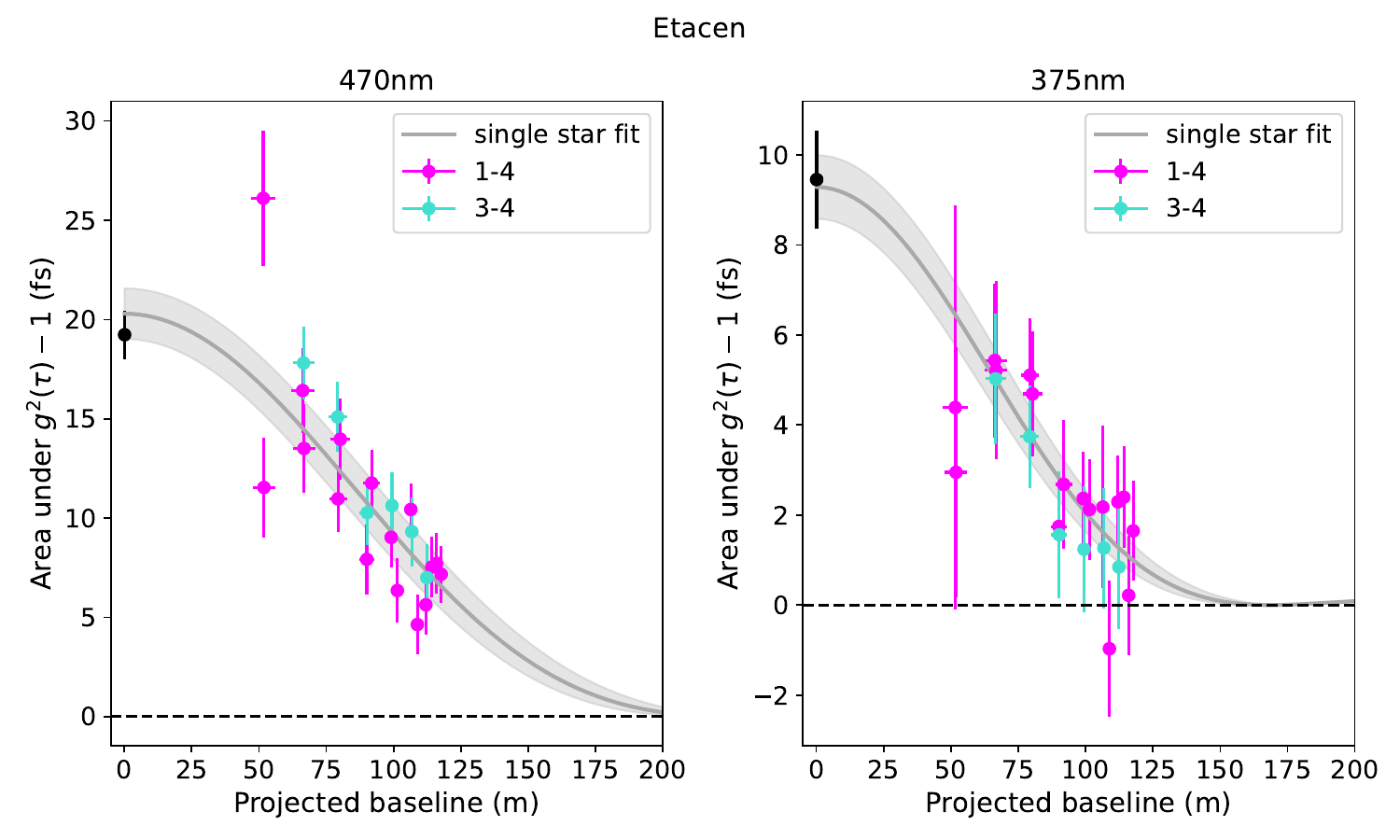}
    \caption{Obtained squared visibility as a function of the projected baseline (m). Upper: Spatial coherence of Mimosa. Lower: Spatial coherence of Eta Centauri. Left: $470\,$nm filter. Right: $375\,$nm filter. The data points resulting from different telescope combinations are presented in different colours. The black data point displays the spatial coherence value at zero baseline. The gray curve represents the fit of the limb darkening model and is referred to as the single star fit.}
    \label{fig:sc_mimosa_etacen}
\end{figure}

\begin{figure}
    \includegraphics[width=\columnwidth]{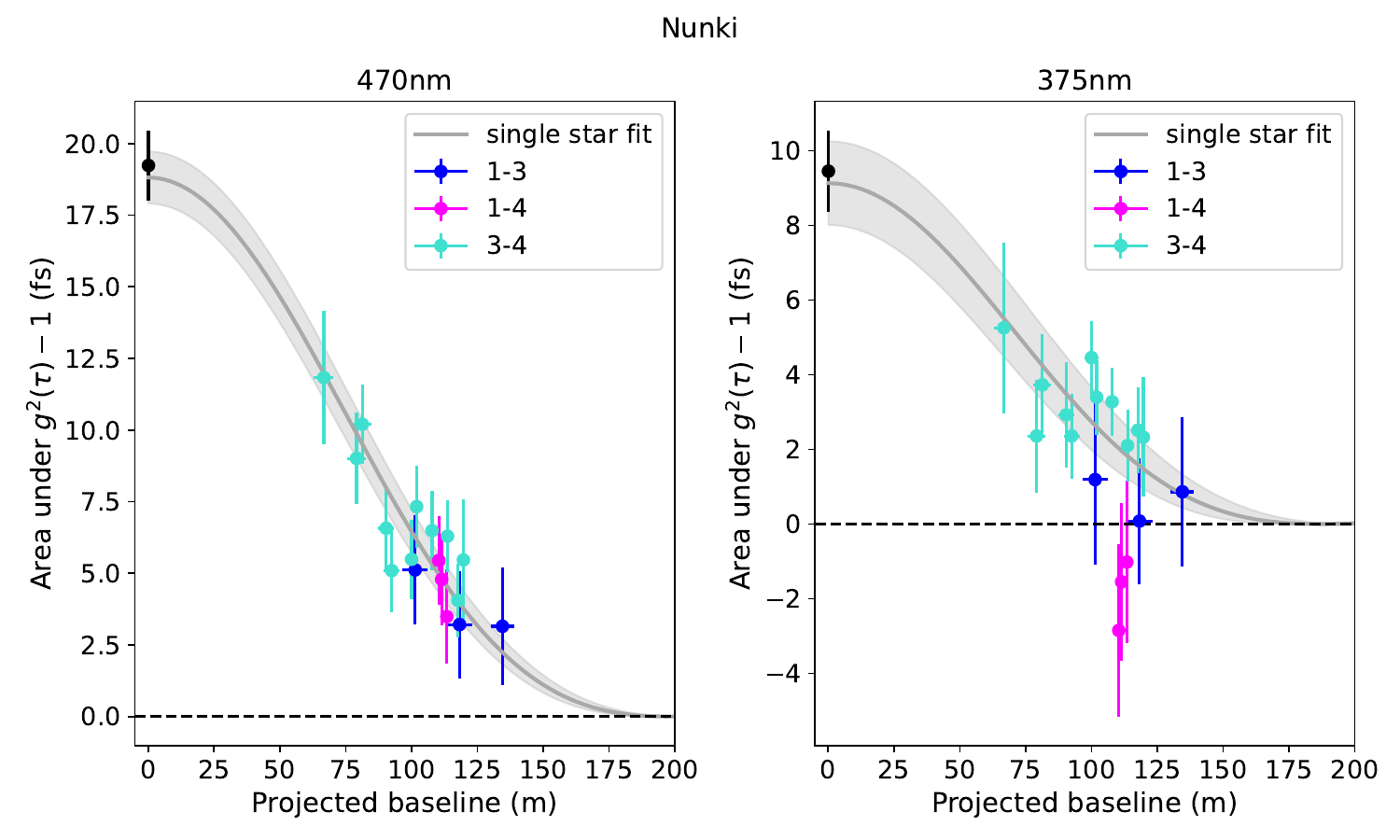}
    \includegraphics[width=\columnwidth]{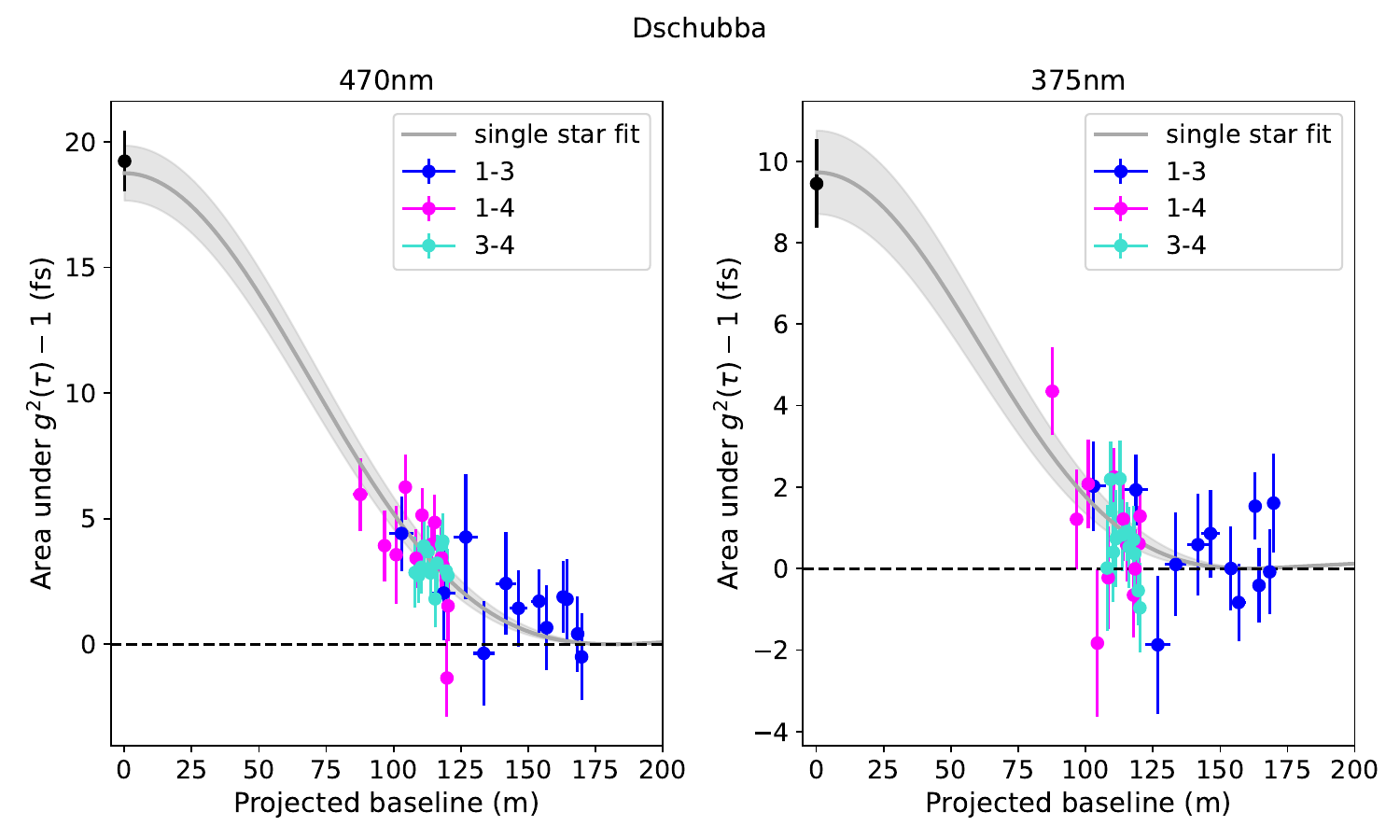}
    \caption{Obtained squared visibility as a function of the projected baseline (m). Upper: Spatial coherence of Nunki. Lower: Spatial coherence of Dschubba. Left: $470\,$nm filter, right: $375\,$nm filter. The data points resulting from different telescope combinations are presented in different colours. The black data point displays the spatial coherence value at zero baseline. The gray curve represents the fit of the limb darkening model and is referred to as the single star fit.}
    \label{fig:sc_nunki_dschubba}
\end{figure}

We present the angular diameters of our measurements in April/May 2023.
All important parameters, which are necessary to compute the diameter and are mentioned in Sect.~\ref{sec:analysis}, are listed in Table \ref{table:parameters}. The table also includes the spectral type of the targets and the reference diameter. The final angular diameters for UD and LD model of our targets are presented in Table \ref{table:angdia}. The squared visibility curves are shown in Fig.~\ref{fig:sc_mimosa_etacen} and Fig.~\ref{fig:sc_nunki_dschubba}. The gray curve represents the LD model which is referred to as the single star fit. As one can see in Table \ref{table:angdia} the UD model underestimates the angular diameter compared to the LD model by $\approx 3\,$\%. If LD is corrected for properly, the LD angular diameter of one star should be the same at different wavelengths, which is not the case in our particular situation. Whilst the $\theta_{UD}$ results for Mimosa and Eta Centauri for both colours are consistent within uncertainties, the results for Nunki and Dschubba are significantly different beyond their error. The cause of these discrepancies is yet unclear. They may either point to additional systematic uncertainties, which we haven't considered, or to a more complex star model (rapid rotator, gravity darkening), where the stars' physical sizes differentiate at different wavelengths. As for Nunki, which is indeed a rapid rotator \citep{Nunki_rot_vel}, the reference value of $0.68\,$mas was measured at infrared wavelength which would be consistent with the assumption that the star is measured larger at larger wavelengths.

For analyzing Dschubba it is important to know its binary characteristics. The two components of the spectroscopic binary are separated by $0.2\,$arcsec and have a magnitude of $2.2$ and $4.6$ \citep{DschubbaMag}. The system has an orbital period of $10.5\,$years and the orbit is elliptical \citep{DschubbaOrbit}. The two stars in the binary are too far apart to be resolved by the H.E.S.S.\ telescopes. A wobbling in the $g^{(2)}$ function caused by the separation of the two stars is expected on the order of $< 1\,$m, which is averaged out by the dish diameter of the single telescopes.\citep{Nitu} 
Furthermore, the difference in magnitude indicates that the secondary has less intensity by a factor of 10. Because of these properties, we decided to present the data with a single star fit as shown in Fig.~\ref{fig:sc_nunki_dschubba} and interpret the resulting angular diameter as the diameter of the primary. The reduced $\chi^2$ for the single star fit for both colours is $0.92$, which is consistent with the single star model.

\begin{table*}
\caption{Stellar targets and their characteristic parameters. Spectral types and B magnitudes are taken from SIMBAD. The reference angular diameters are recovered from NSII \protect\citep{HanburyBrown1974}, where the stars were measured at a wavelength of $443\,$nm, and from \protect\cite{underhill1979effective}. $T_{\text{eff}}$ is the effective temperature of the star and log(g) is the surface gravity. These are parameters needed to derive the LD coefficient $u_\lambda$. $^a$ values are from \protect\cite{Mimosa}, $^b$ values are from \protect\cite{EtaCen}, $^c$ values are from \protect\cite{Nunki}, $^d$ values are from \protect\cite{Dschubba}.}
\centering
\begin{tabular}{ c  c  c  c  c  c  c  c }
 \hline
 \noalign{\smallskip}
Name & \parbox[c]{1cm}{Spectral \\ Type} & \parbox[c]{1cm}{B \\ (mag)} & \parbox[c]{1cm}{Reference \\ $\theta_{\text{UD}}$(mas)} & Source & $T_{\text{eff}}$ (K) & \parbox[c]{1cm}{log(g) \\ $\text{cm}\,{\text{s}^{-2}}$} & $u_{\lambda}$ \\ [0.5ex] 
\noalign{\smallskip}
 \hline\hline
 \noalign{\smallskip}
 Mimosa & B1IV & $1.02$ & $0.702 \pm 0.022$ & NSII & $27000 \pm 1000 ^a$ & $3.6 \pm 0.1 ^a$ & $0.37$  \\
 Eta Centauri & B2Ve & $2.12$ & $0.47 \pm 0.03$ & NSII & $21000 ^b$ & $3.95 ^b$ & $0.371$ \\
 Nunki & B2.5V & $1.923$ & $0.68$ & Underhill &  $18987 ^c$ & $4.0 ^c$ & $0.385$ \\
 Dschubba & B0.3IV & $2.20$ & $0.45 \pm 0.04$ & NSII & $30900 \pm 309 ^d$ & $3.50 \pm 0.04 ^d$ & $0.373$  \\
 \hline
\end{tabular}
\label{table:parameters}
\end{table*}

\begin{table*}
\caption{Table of observed stars and their measured angular diameters listed with their uncertainties for the UD and LD model and for $470\,$nm and $375\,$nm.}
\centering
\begin{tabular}{ c  c  c  c  c }
 \hline
 \noalign{\smallskip}
Name & \multicolumn{2}{c}{$375\,$nm} & \multicolumn{2}{c}{$470\,$nm} \\
 & Measured $\theta_{\text{UD}}$ & Measured $\theta_{\text{LD}}$ & Measured $\theta_{\text{UD}}$ & Measured $\theta_{\text{LD}}$\\ [0.5ex] 
\hline\hline
\noalign{\smallskip}
Mimosa & $0.676\pm 0.017$ & $0.699\pm 0.018$ & $0.654\pm 0.016$ & $0.675\pm 0.016$ \\
Eta Centauri & $0.56\pm 0.02$ & $0.58\pm 0.02$ & $0.53\pm 0.03$ & $0.55\pm 0.03$  \\
Nunki & $0.51\pm 0.03$ & $0.53\pm 0.04$ & $0.61\pm 0.02$ & $0.63\pm0.02$ \\
Dschubba & $0.59\pm 0.02$ & $0.61\pm 0.02$ & $0.66\pm 0.01$ & $0.68\pm0.01$ \\
 \hline
\end{tabular}
\label{table:angdia}
\end{table*}

%\begin{figure}
 %   \includegraphics[width=\columnwidth]{./images/SC_Etacen.pdf}
%    \caption{The Spatial coherence (fs) of Eta Centauri is shown as a function of the projected baseline (m). Left: $470\,$nm filter, right: $375\,$nm filter. The data points resulting from different telescope combinations are presented in different colours. The black data point displays the spatial coherence value at zero baseline. The gray curve represents the fit of the limb darkening model and is referred to as the single star fit.}
%    \label{fig:sc_etacen}
%\end{figure}

%\begin{figure}
%    \includegraphics[width=\columnwidth]{./images/SC_Dschubba.pdf}
%    \caption{The Spatial coherence (fs) of Dschubba is shown as a function of the projected baseline (m). Left: $470\,$nm filter, right: $375\,$nm filter. The data points resulting from different telescope combinations are presented in different colours. The black data point displays the spatial coherence value at zero baseline. The gray curve represents the fit of the limb darkening model and is referred to as the single star fit.}
%    \label{fig:sc_dschubba}
%\end{figure}

\subsection{Data comparison Nunki} \label{Nunki}
\begin{figure}
\centering
    \includegraphics[width=0.8\columnwidth]{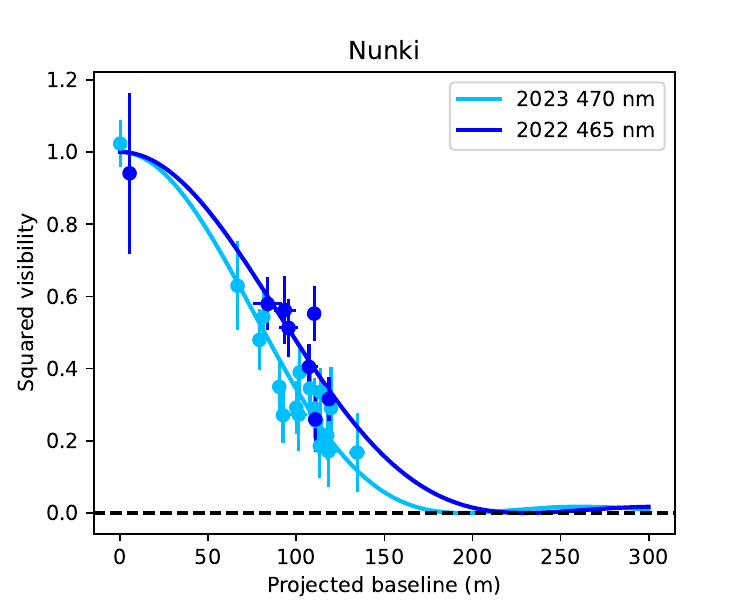}
    \caption{Squared visibility of Nunki is shown as a function of the projected baseline (m). The light blue data represents the results of the 2023 campaign and the dark blue data the 2022 campaign. The data was normalized to one to make the visual comparison easier.}
    \label{fig:Nunki_compare}
\end{figure}

Here we compare the results for the star Nunki, where the reference angular diameter is $0.68\,$mas \citep{underhill1979effective}. Our first measurements from April 2022 resulted in a diameter of $(0.52 \pm 0.07)\,$mas \citep{Zmija_2023}. The optics consisted of an interference filter with $465\,$nm CWL and $2\,$nm width. Further differences in our optical setup to the measurement campaign 2023 can be found in Sect.~\ref{sec:meas_setup}. The angular diameter of the UD model which we determined from the measurements of 2023 is $(0.61\pm 0.02)\,$mas for the interference filter with $470\,$nm CWL and $10\,$nm width. For comparison with the data of 2022 we only take a closer look at the $470\,$nm filter since the CWL are comparable. Figure \ref{fig:Nunki_compare} shows the normalized squared visibility curves of both campaigns. Another aspect, which has to be mentioned is the observation time of the star. In 2022 measurements of Nunki were taken for 11:59 hours, whereas in 2023 it was only 7:25 hours. The results are consistent within $1.2$ standard deviations, where the uncertainty in 2023 was strongly reduced compared to 2022, despite less observation time. The reasons are the (partial) simultaneous operation of 3 telescopes instead of only 2 (which result in 3 simultaneous measurements instead of only one), and the more precise determination of the zero baseline amplitude.

%###############################
\section{Conclusion and outlook}\label{sec:conclusion}
%###############################
In this publication, we have presented our results of the 2023 intensity interferometry campaign at the H.E.S.S.\ telescopes. Our two campaigns of 2022 and 2023 demonstrate how well IACTs are suited for II during full moon periods, when they are not able to perform gamma-ray observations. 

Several goals for updating our setup were met. To gain more telescope baselines three telescopes were equipped with our setup. The bandwidth of the interference filters was increased from $2\,$nm to $10\,$nm and the setup was updated to measure in two colours ($375\,$nm and $470\,$nm) simultaneously. We measured the diameters of 4 stars, not only with the UD model but also taking the LD model into account. The precision of the extracted diameters is at a few percent level within a reasonable measurement time. The results for Mimosa and Eta Centauri between the two colours are consistent within uncertainties. Comparing the resulting angular diameters of Nunki, they get smaller with shorter wavelengths, which is the same case for the results of Dschubba. This effect might be attributed to a more intricate stellar model. H.E.S.S.\ is the first IACT to perform II with two colours simultaneously. This marks substantial progress towards implementing II with CTAO, as this approach might eventually set the standard for these measurements and promises enhancements in signal-to-noise ratio. 

In the future, all four H.E.S.S.\ Phase I telescopes will be equipped with an II setup to supply a broad coverage in the two dimensional baseline map ("uv plane"). Furthermore, the mechanical and computational setup will be updated in order to be able to execute observations remotely. A method to restrict the value of the zero baseline amplitude would be to have a calibration source or star, which produces strong correlation signals due to its brightness and covers a large baseline range for good restriction with the fit model. This is done by MAGIC and described in \cite{Abe_2024}. For the H.E.S.S. telescopes the ideal target would be the star Mimosa. Future targets will involve fainter single stars to determine the sensitivity of our system. Not only single stars in the southern hemisphere will be observed but also binary star systems with various characteristics. 

%##########################
\section*{Acknowledgements}
%##########################
%The Acknowledgements section is not numbered. Here you can thank helpful colleagues, acknowledge funding agencies, telescopes and facilities used etc. Try to keep it short.
We thank the H.E.S.S. collaboration for reviewing this work and for allowing us to use their telescopes for our measurement campaign.
We also thank the H.E.S.S. local crew for their support on site and the operations team for their help online.
This work was supported with a grant by the Deutsche Forschungsgemeinschaft (\lq Optical intensity interferometry with the H.E.S.S. gamma-ray telescopes\rq - FU 1093/3-1).

%#############################################################################
\section*{Data Availability}
The data underlying this article will be shared on reasonable request to the corresponding author. Correlation histograms are available in time-intervals of $3.436\,$s. Due to the large size of the digitized waveforms in excess of several TB, the raw data cannot be made available online.



\bibliographystyle{mnras}
\bibliography{cites} % if your bibtex file is called cites.bib

\begin{thebibliography}{}
\makeatletter
\relax
\def\mn@urlcharsother{\let\do\@makeother \do\$\do\&\do\#\do\^\do\_\do\%\do\~}
\def\mn@doi{\begingroup\mn@urlcharsother \@ifnextchar [ {\mn@doi@} {\mn@doi@[]}}
\def\mn@doi@[#1]#2{\def\@tempa{#1}\ifx\@tempa\@empty \href {http://dx.doi.org/#2} {doi:#2}\else \href {http://dx.doi.org/#2} {#1}\fi \endgroup}
\def\mn@eprint#1#2{\mn@eprint@#1:#2::\@nil}
\def\mn@eprint@arXiv#1{\href {http://arxiv.org/abs/#1} {{\tt arXiv:#1}}}
\def\mn@eprint@dblp#1{\href {http://dblp.uni-trier.de/rec/bibtex/#1.xml} {dblp:#1}}
\def\mn@eprint@#1:#2:#3:#4\@nil{\def\@tempa {#1}\def\@tempb {#2}\def\@tempc {#3}\ifx \@tempc \@empty \let \@tempc \@tempb \let \@tempb \@tempa \fi \ifx \@tempb \@empty \def\@tempb {arXiv}\fi \@ifundefined {mn@eprint@\@tempb}{\@tempb:\@tempc}{\expandafter \expandafter \csname mn@eprint@\@tempb\endcsname \expandafter{\@tempc}}}

\bibitem[\protect\citeauthoryear{Abe et~al.,}{Abe et~al.}{2024}]{Abe_2024}
Abe S.,  et~al., 2024, \mn@doi [Monthly Notices of the Royal Astronomical Society] {10.1093/mnras/stae697}, 529, 4387–4404

\bibitem[\protect\citeauthoryear{Abeysekara et~al.,}{Abeysekara et~al.}{2020}]{abeysekara2020demonstration}
Abeysekara A.,  et~al., 2020, Nature Astronomy, 4, 1164

\bibitem[\protect\citeauthoryear{Abt, Levato  \& Grosso}{Abt et~al.}{2002}]{Nunki_rot_vel}
Abt H.~A.,  Levato H.,   Grosso M.,  2002, \mn@doi [The Astrophysical Journal] {10.1086/340590}, 573, 359

\bibitem[\protect\citeauthoryear{Acciari et~al.,}{Acciari et~al.}{2020}]{acciari2020optical}
Acciari V.,  et~al., 2020, Monthly Notices of the Royal Astronomical Society, 491, 1540

\bibitem[\protect\citeauthoryear{Acharyya et~al.,}{Acharyya et~al.}{2024}]{acharyya2024}
Acharyya A.,  et~al., 2024, An Angular Diameter Measurement of $\beta$ UMa via Stellar Intensity Interferometry with the VERITAS Observatory (\mn@eprint {arXiv} {2401.01853}), \url {https://arxiv.org/abs/2401.01853}

\bibitem[\protect\citeauthoryear{{Adnacom}}{{Adnacom}}{2023}]{pcie}
{Adnacom} 2023, {R34 PCIe board}, \url {https://adnacom.com/r34/}

\bibitem[\protect\citeauthoryear{{Arcos}, {Kanaan}, {Ch{\'a}vez}, {Vanzi}, {Araya}  \& {Cur{\'e}}}{{Arcos} et~al.}{2018a}]{EtaCen}
{Arcos} C.,  {Kanaan} S.,  {Ch{\'a}vez} J.,  {Vanzi} L.,  {Araya} I.,   {Cur{\'e}} M.,  2018a, \mn@doi [\mnras] {10.1093/mnras/stx3075}, \href {https://ui.adsabs.harvard.edu/abs/2018MNRAS.474.5287A} {474, 5287}

\bibitem[\protect\citeauthoryear{Arcos, Kanaan, Chávez, Vanzi, Araya  \& Curé}{Arcos et~al.}{2018b}]{Dschubba}
Arcos C.,  Kanaan S.,  Chávez J.,  Vanzi L.,  Araya I.,   Curé M.,  2018b, \mn@doi [Monthly Notices of the Royal Astronomical Society] {10.1093/mnras/stx3075}, 474, 5287

\bibitem[\protect\citeauthoryear{{Claret, A.} \& {Bloemen, S.}}{{Claret, A.} \& {Bloemen, S.}}{2011}]{LD}
{Claret, A.} {Bloemen, S.} 2011, \mn@doi [AandA] {10.1051/0004-6361/201116451}, 529, A75

\bibitem[\protect\citeauthoryear{Cohen, Kuhn, Gagné, Jensen  \& Miller}{Cohen et~al.}{2008}]{Mimosa}
Cohen D.~H.,  Kuhn M.~A.,  Gagné M.,  Jensen E. L.~N.,   Miller N.~A.,  2008, \mn@doi [Monthly Notices of the Royal Astronomical Society] {10.1111/j.1365-2966.2008.13176.x}, 386, 1855

\bibitem[\protect\citeauthoryear{Cortina et~al.,}{Cortina et~al.}{2022}]{cortina2022first}
Cortina J.,  et~al., 2022, in Optical and Infrared Interferometry and Imaging VIII. pp 127--141

\bibitem[\protect\citeauthoryear{Dravins}{Dravins}{2016}]{dravins2016intensity}
Dravins D.,  2016, in Optical and Infrared Interferometry and Imaging V. pp 128--139

\bibitem[\protect\citeauthoryear{Dravins, LeBohec, Jensen  \& Nuñez}{Dravins et~al.}{2013}]{Dravins2013}
Dravins D.,  LeBohec S.,  Jensen H.,   Nuñez P.~D.,  2013, \mn@doi [Astroparticle Physics] {https://doi.org/10.1016/j.astropartphys.2012.04.017}, 43, 331

\bibitem[\protect\citeauthoryear{Ferreira, Bachelard, Guerin, Kaiser  \& Fouché}{Ferreira et~al.}{2020}]{Siegert_Ferreira}
Ferreira D.,  Bachelard R.,  Guerin W.,  Kaiser R.,   Fouché M.,  2020, \mn@doi [American Journal of Physics] {10.1119/10.0001630}, 88, 831

\bibitem[\protect\citeauthoryear{Haguenauer et~al.,}{Haguenauer et~al.}{2010}]{haguenauer2010very}
Haguenauer P.,  et~al., 2010, in Optical and Infrared Interferometry II. pp 53--63

\bibitem[\protect\citeauthoryear{{Hamamatsu, Inc}}{{Hamamatsu, Inc}}{2023}]{Hamamatsu}
{Hamamatsu, Inc} 2023, {Hamamatsu Photomultiplier Tubes}, \url {https://www.hamamatsu.com/content/dam/hamamatsu-photonics/sites/documents/99_SALES_LIBRARY/etd/R11265U_H11934_TPMH1336E.pdf}

\bibitem[\protect\citeauthoryear{Hanbury~Brown}{Hanbury~Brown}{1974}]{HanburyBrown1974}
Hanbury~Brown R.,  1974, The intensity interferometer; its application to astronomy.
Taylor \& Francis, Halsted Press

\bibitem[\protect\citeauthoryear{Hanbury~Brown, Davis  \& Allen}{Hanbury~Brown et~al.}{1967}]{brown1967stellar}
Hanbury~Brown R.,  Davis J.,   Allen L.~R.,  1967, Monthly Notices of the Royal Astronomical Society, 137, 375

\bibitem[\protect\citeauthoryear{Hanbury~Brown, Davis  \& Allen}{Hanbury~Brown et~al.}{1974}]{HBT_32}
Hanbury~Brown R.,  Davis J.,   Allen L.~R.,  1974, \mn@doi [Monthly Notices of the Royal Astronomical Society] {10.1093/mnras/167.1.121}, 167, 121

\bibitem[\protect\citeauthoryear{Hofmann}{Hofmann}{2012}]{HESStel}
Hofmann W.,  2012, High Energy Stereoscopic System, \url {https://www.mpi-hd.mpg.de/HESS/pages/about/telescopes/}

\bibitem[\protect\citeauthoryear{Kieda}{Kieda}{2022}]{kieda2022performance}
Kieda D.~B.,  2022, in Optical and Infrared Interferometry and Imaging VIII. pp 142--156

\bibitem[\protect\citeauthoryear{Klepser}{Klepser}{2012}]{HESSpic}
Klepser 2012, {High Energy Stereoscopic System}, \url {https://en.wikipedia.org/wiki/High_Energy_Stereoscopic_System}

\bibitem[\protect\citeauthoryear{{Laser Components, Inc}}{{Laser Components, Inc}}{2023}]{Lasercomponents}
{Laser Components, Inc} 2023, {Laser Components}, \url {https://www.lasercomponents.com/de/produkt/standard-bandpassfilter/}

\bibitem[\protect\citeauthoryear{LeBohec \& Holder}{LeBohec \& Holder}{2006}]{LeBohec2006}
LeBohec S.,  Holder J.,  2006, \mn@doi [The Astrophysical Journal] {10.1086/506379}, 649, 399

\bibitem[\protect\citeauthoryear{Mandel \& Wolf}{Mandel \& Wolf}{1995}]{Mandel1995}
Mandel L.,  Wolf E.,  1995, Optical Coherence and Quantum Optics.
Cambridge University Press, \mn@doi{10.1017/CBO9781139644105}

\bibitem[\protect\citeauthoryear{Mason, Wycoff, Hartkopf, Douglass  \& Worley}{Mason et~al.}{2001}]{DschubbaMag}
Mason B.~D.,  Wycoff G.~L.,  Hartkopf W.~I.,  Douglass G.~G.,   Worley C.~E.,  2001, \mn@doi [The Astronomical Journal] {10.1086/323920}, 122, 3466

\bibitem[\protect\citeauthoryear{Miroshnichenko et~al.,}{Miroshnichenko et~al.}{2013}]{DschubbaOrbit}
Miroshnichenko A.~S.,  et~al., 2013, \mn@doi [The Astrophysical Journal] {10.1088/0004-637x/766/2/119}, 766, 119

\bibitem[\protect\citeauthoryear{Rai, Basak  \& Saha}{Rai et~al.}{2021}]{Nitu}
Rai K.~N.,  Basak S.,   Saha P.,  2021, \mn@doi [Monthly Notices of the Royal Astronomical Society] {10.1093/mnras/stab2391}, 507, 2813

\bibitem[\protect\citeauthoryear{Siegert}{Siegert}{1943}]{siegert1943fluctuations}
Siegert A. . M. I. o. T. R.~L.,  1943, On the Fluctuations in Signals Returned by Many Independently Moving Scatterers.
Report (Massachusetts Institute of Technology. Radiation Laboratory), Radiation Laboratory, Massachusetts Institute of Technology, \url {https://books.google.de/books?id=5PzqGwAACAAJ}

\bibitem[\protect\citeauthoryear{{Spectrum Instrumentation}}{{Spectrum Instrumentation}}{2023}]{specint}
{Spectrum Instrumentation} 2023, {M4i.2211-x8 digitizer}, \url {https://spectrum-instrumentation.com/products/details/M4i2211-x8.php}

\bibitem[\protect\citeauthoryear{{Thorlabs, Inc}}{{Thorlabs, Inc}}{2023}]{Thorlabs}
{Thorlabs, Inc} 2023, {Thorlabs}, \url {https://www.thorlabs.com/navigation.cfm?guide_id=2264}

\bibitem[\protect\citeauthoryear{Underhill, Divan, Prevot-Burnichon  \& Doazan}{Underhill et~al.}{1979a}]{underhill1979effective}
Underhill A.,  Divan L.,  Prevot-Burnichon M.-L.,   Doazan V.,  1979a, Monthly Notices of the Royal Astronomical Society, 189, 601

\bibitem[\protect\citeauthoryear{Underhill, Divan, Prévot-Burnichon  \& Doazan}{Underhill et~al.}{1979b}]{Nunki}
Underhill A.~B.,  Divan L.,  Prévot-Burnichon M.-L.,   Doazan V.,  1979b, \mn@doi [Monthly Notices of the Royal Astronomical Society] {10.1093/mnras/189.3.601}, 189, 601

\bibitem[\protect\citeauthoryear{Zmija, Vogel, Anton, Malyshev, Michel, Zink  \& Funk}{Zmija et~al.}{2021}]{10.1093/mnras/stab3058}
Zmija A.,  Vogel N.,  Anton G.,  Malyshev D.,  Michel T.,  Zink A.,   Funk S.,  2021, \mn@doi [Monthly Notices of the Royal Astronomical Society] {10.1093/mnras/stab3058}, 509, 3113

\bibitem[\protect\citeauthoryear{Zmija, Vogel, Wohlleben, Anton, Zink  \& Funk}{Zmija et~al.}{2023}]{Zmija_2023}
Zmija A.,  Vogel N.,  Wohlleben F.,  Anton G.,  Zink A.,   Funk S.,  2023, \mn@doi [Monthly Notices of the Royal Astronomical Society] {10.1093/mnras/stad3676}, 527, 12243–12252

\bibitem[\protect\citeauthoryear{ten Brummelaar et~al.,}{ten Brummelaar et~al.}{2005}]{CHARA_description}
ten Brummelaar T.~A.,  et~al., 2005, \mn@doi [The Astrophysical Journal] {10.1086/430729}, 628, 453–465

\makeatother
\end{thebibliography}

%\begin{thebibliography}{99}
%\end{thebibliography}

%%%%%%%%%%%%%%%%%%%%%%%%%%%%%%%%%%%%%%%%%%%%%%%%%%

%%%%%%%%%%%%%%%%% APPENDICES %%%%%%%%%%%%%%%%%%%%%

% \appendix

% \section{Some extra material}

% If you want to present additional material which would interrupt the flow of the main paper,
% it can be placed in an Appendix which appears after the list of references.

%%%%%%%%%%%%%%%%%%%%%%%%%%%%%%%%%%%%%%%%%%%%%%%%%%

% Don't change these lines
\bsp	% typesetting comment
\label{lastpage}
\end{document}